\documentclass[%
twocolumn,
nofootinbib,
amsmath,amssymb,
nobalancelastpage,
aps,
prl,
floatfix,
longbibliography,
]{revtex4-2}

\usepackage{graphicx}
\usepackage{dcolumn}
\usepackage{bm}
\usepackage{siunitx}
\usepackage{physics}
\AtBeginDocument{\RenewCommandCopy\qty\SI}

\usepackage{easyReview}
\usepackage{slashed}
\usepackage{mathtools}
\usepackage{xspace}
\usepackage{comment}
\usepackage{braket}
\usepackage{float}
\usepackage{fontawesome}
\usepackage{booktabs}
\usepackage{anyfontsize}
\usepackage{listings}
\usepackage{inconsolata}
\usepackage{multirow}
\usepackage{tikz}
\usepackage{xpatch}
\xpatchbibdriver{misc}
  {\printtext[parens]{\usebibmacro{date}}}
  {\iffieldundef{year}
    {}
    {\printtext[parens]{\usebibmacro{date}}}}
  {}
  {\typeout{Removed date because none provided}}

\tikzset{every picture/.style={line width=0.75pt}}
\usetikzlibrary{calc}
\usetikzlibrary{decorations.pathmorphing}

\definecolor{linkcolor}{rgb}{0.0, 0.28, 0.67}
\definecolor{userInputColor}{HTML}{4287f5}
\definecolor{userCallColor}{HTML}{fcba03}
\definecolor{codeCallColor}{HTML}{039c33}
\usepackage[dvipsnames]{xcolor}

\usepackage[
    colorlinks=true,
    urlcolor=linkcolor,
    anchorcolor=linkcolor,
    citecolor=linkcolor,
    filecolor=linkcolor,
    linkcolor=linkcolor,
    menucolor=linkcolor,
    linktocpage=true,
    pdfproducer=medialab,
    pdfa=true
]{hyperref}

\DeclareSIUnit\erg{erg}
\DeclareSIUnit\year{yr}
\DeclareSIUnit\jansky{Jy}
\DeclareSIUnit\dex{dex}
\DeclareSIUnit\deg{deg}
\DeclareSIUnit\angstrom{\text{Å}}
\DeclareSIUnit\eV{e\kern-.05em V}
\DeclareSIUnit \parsec {pc}
\DeclareSIUnit \AU {AU}
\DeclareSIUnit \gauss {G}

\DeclareSIUnit[]\msol
{\text{\ensuremath{M_{\odot}}}}
\DeclareSIUnit\beam{beam}
\graphicspath{ {../images/} }

\defcitealias{Baker:2025bnz}{Paper~II}
\usepackage[normalem]{ulem}

\begin{document}

\title{Radio Emission from High-Frequency Gravitational Wave Point Sources}
\author{Ethan Baker}
\email{ebaker@bu.edu}
\thanks{ORCID: \href{https://orcid.org/0000-0002-0520-4235}{0000-0002-0520-4235}}
\affiliation{Physics Department, Boston University, Boston, MA 02215, USA}

\author{Hongwan Liu}
\email{hongwan@bu.edu}
\thanks{ORCID: \href{https://orcid.org/0000-0003-2486-0681}{0000-0003-2486-0681}}
\affiliation{Physics Department, Boston University, Boston, MA 02215, USA}

\date{\today}

\begin{abstract}
    High-frequency gravitational waves (HFGWs) in the MHz to GHz regime can convert into radio photons in the presence of astrophysical magnetic fields through the inverse Gertsenshtein effect. We show that existing radio telescopes like CHIME and FAST are excellent tools for detecting HFGW sources, significantly outperforming many existing experiments at detecting primordial black hole (PBH) mergers, the most realistic sources of transient HFGWs. Radio telescopes are also uniquely sensitive to sources of monochromatic HFGW emission, such as ultralight boson clouds formed through superradiance around PBHs, and are likely to have excellent sensitivity to generic sources of detectable HFGWs. 
\end{abstract}

\maketitle

\paragraph*{\textbf{Introduction.}}
Gravitational wave (GW) interferometers (e.g.\ LIGO \cite{LIGOScientific:2007fwp,LIGOScientific:2014pky,LIGOScientific:2020ibl}, VIRGO \cite{VIRGO:2012dcp}, and KAGRA \cite{KAGRA:2020cvd}) and pulsar timing arrays (e.g.\ CPTA \cite{Xu:2023wog}, EPTA \cite{EPTA:2016ndq}, InPTA \cite{Tarafdar:2022toaa}, MeerTime \cite{Bailes:2018azh}, NANOGrav \cite{Brazier:2019mmu}, and PPTA \cite{Hobbs:2013aka}) have revolutionized our understanding of black holes and other compact objects by detecting GWs with frequencies in the \unit{\nano\hertz} to \unit{\kilo\hertz} range \cite{LIGOScientific:2017adfa,NANOGrav:2023gor,NANOGrav:2023hvm,LIGOScientific:2025kry,LIGOScientific:2025brd,LIGOScientific:2025wao}. 
The extraordinary success of these experiments has driven significant experimental interest in extending the search for GWs into the \unit{\mega\hertz} to \unit{\giga\hertz} range (see Refs.~\cite{Aggarwal:2020olq,Franciolini:2022htd} for reviews).
Proposals to detect such high-frequency gravitational waves (HFGWs) draw inspiration from another rapidly developing field: experimental searches for axion dark matter. 
In the presence of an external magnetic field, GWs convert into electromagnetic radiation through a Standard-Model process known as the inverse Gertsenshtein effect~\cite{Gertsenshtein:1961,Boccaletti:1970pxw,zel1973electromagnetic}. 
Because of the similarity of this process to the Primakoff effect for axions~\cite{Sikivie:1983ip}, both existing and future axion searches can be repurposed to search for HFGWs~\cite{Ringwald:2020ist,Berlin:2021txa,Domcke:2022rgu,Domcke:2023bat,Capdevilla:2025omb}.
Dedicated proposals such as GravNet---a superconducting-cavity network designed for HFGWs---have also garnered significant support~\cite{Schmieden:2023fzn}. 

A closer look at the possible sources of HFGWs, however, makes the prospects less rosy.
Stochastic or cosmological HFGW backgrounds are tightly constrained by overclosure of the universe, putting them beyond near-term experimental reach (see e.g.\ Refs.~\cite{Domcke:2020yzq,Gupta:2026scx}).
On relatively model-independent grounds, individual HFGW sources in the GHz regime with a reasonable chance of detection must likely lie within the Solar System~\cite{Berlin:2026che}.
At present, the only known detectable Standard Model (SM) sources of transient HFGWs are phase transitions within neutron stars~\cite{Bleau:2026ala}; going beyond the SM, the most promising new-physics sources are mergers of light primordial black holes (PBHs) with masses $\qty{e-13}{\msol}\lesssim m \lesssim \qty{e-6}{\msol}$, which includes the mass range where PBHs could be all of the dark matter \cite{Franciolini:2022htd}. 
In both cases, the HFGW signal is short-lived and its frequency evolves rapidly, making it challenging to detect at repurposed axion experiments, which typically target highly monochromatic signals.

In this \textit{Letter}, we demonstrate that radio telescopes are among the best existing ways to detect realistic HFGW point sources such as PBH mergers, and should be the cornerstone of any concerted search effort.
As GWs from such sources propagate to us, a fraction of the gravitational radiation converts to electromagnetic radiation in astrophysical magnetic fields. 
Unlike previous proposals to use radio telescopes to look for stochastic HFGW backgrounds~\cite{Domcke:2020yzq,Gupta:2026scx}, we show that PBH mergers appear as point-like radio bursts with negligible or even negative dispersion measure (DM), and are detectable if sufficiently bright and within the telescope band. 
We demonstrate that CHIME~\cite{CHIME:2022dwe} and FAST~\cite{zotero-item-3161,zotero-item-3149,zotero-item-3151} can already detect mergers of PBHs with mass $m \sim \qty{e-6}{\msol}$ out to $\sim \qty{1000}{\AU}$, dramatically extending present sensitivity and covering a significant portion of the parameter space targeted by the most optimistic proposed HFGW searches.
Radio facilities can even set leading limits on monochromatic HFGW point sources---such as GW emission from ultralight boson clouds formed by superradiance around PBHs---where cavity-based experiments might be expected to excel, although such sources are considerably less well-motivated physically.
Although we focus on these particular cases, radio telescopes are likely highly sensitive to any generic source of detectable HFGWs like those considered in Ref.~\cite{Berlin:2026che} through the same mechanisms that we describe here. 
Throughout this work, we use natural units where $\hbar=c=1$, except in radio-astronomy contexts where we use units standard in that field.

\paragraph*{\textbf{Inverse Gertsenshtein Effect.}}

In the presence of a background magnetic field, gravitons and photons mix~\cite{Gertsenshtein:1961,Boccaletti:1970pxw,zel1973electromagnetic}.
The probability of a graviton converting into a photon $P_{h\to \gamma}$ can be computed by solving the coupled equations of motion for GWs and the electromagnetic field. 
Explicitly, over a distance $\ell$ in a region with uniform magnetic field $B$~\cite{Domcke:2020yzq,Dunsky:2025pvd}, 
\begin{equation}
    P^{(1)}_{h \to \gamma} (\ell) = \frac{\mu}{(1+\mu)^2} \kappa^2 B^2\ell_{\rm osc}^2 \sin^2\left(\frac{\ell}{\ell_{\rm osc}}\right)\, ,
    \label{eq:prob_one_patch}
\end{equation}
where $\kappa^2 = 16\pi G_N$, $\ell_{\rm osc} \equiv 2 [\omega_{\rm GW}^2(1-\mu)^2 + \kappa^2B^2]^{-1/2}$ is the characteristic oscillation length-scale, $\mu \equiv \sqrt{1-\omega_{\rm pl}^2/\omega_{\rm GW}^2}\approx 1-\omega_{\rm pl}^2/2\omega_{\rm GW}^2$ is the electromagnetic refractive index, $\omega_{\rm GW} \equiv 2 \pi f_{\rm GW}$ is the photon and graviton frequency, and $\omega_{\rm pl}^2\approx 4\pi \alpha_{\rm EM}n_e/m_e$ is the plasma mass~\cite{Domcke:2020yzq}. 
Here, $G_N$ is Newton's gravitational constant, $\alpha_{\rm EM}$ is the fine-structure constant, $n_e$ is the free-electron number density, and $m_e$ is the electron mass.
For the weak astrophysical magnetic fields considered here, the first term in $\ell_{\rm osc}$ dominates, so $\ell_{\rm osc} \approx 4 \omega_{\rm GW}/\omega_{\rm pl}^2$. 

Eq.~\eqref{eq:prob_one_patch} is the solution for traversing a region with uniform magnetic field.
In reality, as the GW propagates to the observer, the magnetic field changes over a length scale given by the coherence length, $\ell_c$.
The total conversion probability $P_{h \to \gamma}(d)$ from a distance $d$ to Earth is then approximately obtained by summing over contributions from patches $i$ with coherence length $\ell_{c,i}$~\cite{Domcke:2020yzq}, i.e.
\begin{equation}
    P_{h \to \gamma}(d) \approx \sum_{i=1}^{N_{\rm tot}-1} \frac{\mu_i}{(1+\mu_i)^2} \kappa^2 B_i^2\ell_{{\rm osc}, i}^2 \sin^2\left(\frac{\ell_{c, i}}{\ell_{{\rm osc}, i}}\right)\, ,
\end{equation}
where $N_{\rm tot}$ is the total number of patches considered. 
In practice, starting at distance $d$ from Earth, we compute $B$, $\ell_c$, and $n_e$ from our model (below), add the single-patch conversion probability to the running total, step toward Earth by $\ell_c$, and repeat until the remaining distance falls below $\ell_c$.
For the final patch, we compute $P_{h\to \gamma}^{(1)}$ using Eq.~(34) of Ref.~\cite{Domcke:2020yzq}, which accounts for conversions in a patch smaller than $\ell_c$.
In general, this contribution is negligible except for sources of HFGWs within $\sim r_\oplus$ of the Earth.
Computing $P_{h\to\gamma}$ therefore requires a model for $B$, $\ell_c$, and $n_e$ along the line of sight, which we describe next.

\paragraph*{\textbf{Magnetic Field Model.}}
Since radio telescopes are sensitive to nearby sources, we construct a simplified model of $B$, $\ell_c$, and $n_e$ in the Solar System by log-log interpolating in distance $d$ between the reference values listed in Table~\ref{tab:B_field_model}.
Our model neglects more complicated effects such as the Earth's turbulent field components and the directionality of the solar wind. 

Near Earth, reference values are drawn from satellite measurements, including the Magnetospheric Multiscale Mission~\cite{Burch:2016,xu:2021abc}, ARTEMIS~\cite{angelopoulos:2011,sibeck:2011,Runov:2023abc}, and lunar probes~\cite{Meng:1974abc}.
Within $\sim r_\oplus$ of Earth's surface, we adopt a dipole model of $B \propto r^{-3}$, normalized to \qty{0.25}{\gauss} at Earth's surface~\cite{Finlay:2010lju}, and assume that the atmosphere is neutral at the surface.
At larger distances, we rely on Voyager 1 and 2 data \cite{Gurnett:2019abc} and stellar absorption features that measure $n_e$~\cite{Redfield:2008abc}.

The coherence length $\ell_c$ is less well-characterized, requiring assumptions near Earth and at the Solar System's edge. Since Earth's dipole is globally coherent~\cite{Finlay:2010lju}, we take $\ell_c=r_\oplus$ close to Earth; beyond several hundred \unit{AU}, where no measurements exist, we adopt $\ell_c=\qty{100}{\AU}$, where Voyager 1 observations suggest the magnetic field is highly ordered~\cite{Burlaga:2016abc}.

\begin{table}[t]
    \begin{ruledtabular}
    \begin{tabular}{lccc}
        $d_{\rm ref}$   & $B$ [\unit{\gauss}] & $\ell_c$ & $n_e$ [\unit{\per\centi\meter\cubed}] \\ \colrule
        0               & 0.25~\cite{Finlay:2010lju}              & $r_\oplus$ & 0   \\
        $r_{\oplus}$    & 0.03               & $r_\oplus$ & 3900~\cite{Carpenter:1992abc}\\
        $10r_{\oplus}$  & $3\times 10^{-4}$~\cite{Stawarz:2022abc} & \qty{7.9e-8}{\AU}~\cite{Stawarz:2022abc}& 5.8~\cite{Carpenter:1992abc}\\
        \qty{1}{\AU}    & $5 \times 10^{-5}$~\cite{Goldstein:2005abc} & \qty{0.01}{\AU}~\cite{Goldstein:2005abc}             & 5~\cite{Goldstein:2005abc}\\
        \qty{100}{\AU}  & $5 \times 10^{-6}$~\cite{Gurnett:2019abc} & \qty{0.1}{\AU}~\cite{Fraternale:2022}               & 0.001~\cite{Gurnett:2019abc}\\
        \qty{1000}{\AU} & $5 \times 10^{-6}$~\cite{Burlaga:2016abc} & \qty{100}{\AU}              & 0.1~\cite{Redfield:2008abc}\\
    \end{tabular}
    \end{ruledtabular}
    \caption{Reference values of $B$, $\ell_c$, and $n_e$ at distances $d_{\rm ref}$ from Earth's surface. Within $r_\oplus$ we assume a globally coherent dipole~\cite{Finlay:2010lju}; values at arbitrary distances are obtained by log-log interpolation.}
    \label{tab:B_field_model}
\end{table}

\begin{figure}[t]
    \centering
    \includegraphics[width=\linewidth]{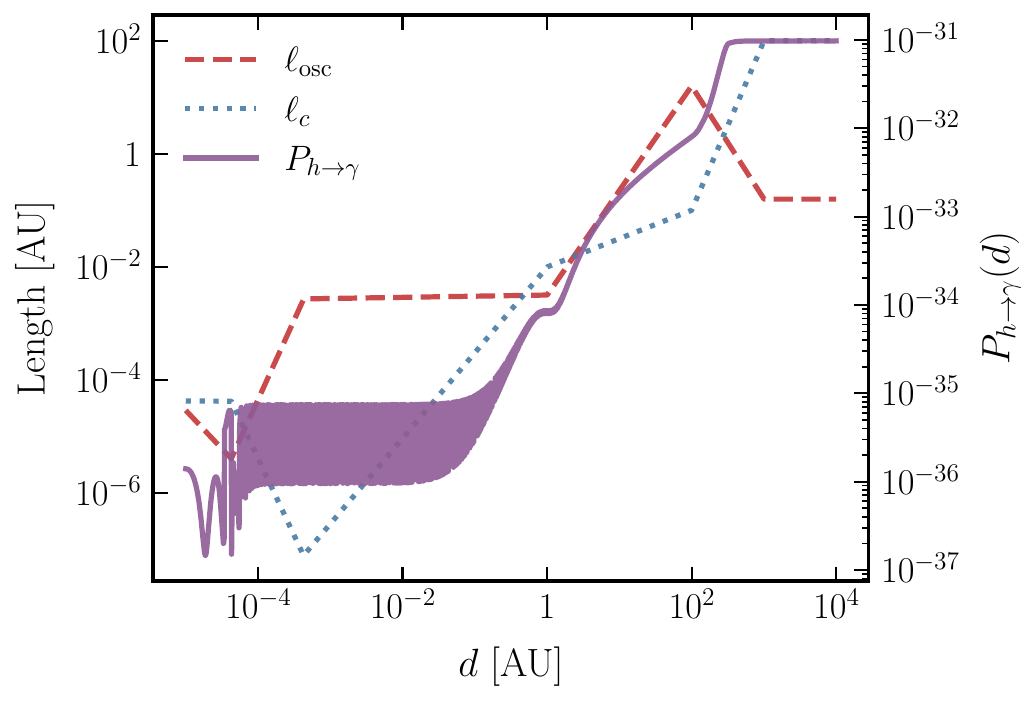}
    \caption{Oscillation length $\ell_{\rm osc}$, coherence length $\ell_c$, and conversion probability $P_{h\to\gamma}(d)$ vs.\ distance $d$ from Earth at $f_{\rm GW}=\qty{1}{\giga\hertz}$.}
    \label{fig:lengths_and_P}
\end{figure}

In Fig.~\ref{fig:lengths_and_P}, we show $\ell_{\rm osc}$ and $\ell_c$ as a function of distance from Earth. Additionally, we plot the total probability of conversion for a source located at a distance $d$ from Earth. 
For distant sources, the probability of conversion is dominated by conversions far from Earth, so $P_{h \to \gamma}(d)$ grows with $d$ before leveling off at $\sim \qty{100}{\AU}$. Near the Earth, $\ell_{\rm osc} \gg \ell_c$, which gives rise to the rapidly oscillatory features in $P_{h \to \gamma}$ for sources located between $\qty{e-4}{\AU}\lesssim d \lesssim \qty{0.1}{\AU}$.

\paragraph*{\textbf{Primordial Black Hole Mergers.}}

With a model for $P_{h \to \gamma}$, we now determine the radio signatures of PBH mergers.
As two PBHs inspiral, they emit a total GW energy per unit frequency~\cite{Maggiore:2007abc}
\begin{equation}
    \frac{dE_{\rm GW}}{df_{\rm GW}} = \frac{\pi}{3 G_N} (G_N m)^{5/3}(2 \pi f_{\rm GW})^{-1/3} \, ,
\end{equation}
where $m$ is the mass of each PBH.
We consider equal-mass mergers in circular orbits for simplicity; mergers with a chirp mass given by $m$ will give parametrically similar results.
The corresponding GW strain at distance $d$ is $h(d)\sim 2(G_Nm)^{5/3}\omega_{\rm GW}^{2/3}/d$ (up to inclination)~\cite{Franciolini:2022htd}, i.e.
\begin{equation}
    \label{eq:pbh_strain}
    h(d) \sim 10^{-19} \left( \frac{m}{\qty{e-7}{\msol}} \right)^{5/3}\!\! \left( \frac{f_{\rm GW}}{\qty{1}{\GHz}} \right)^{2/3} \!\! \left( \frac{\qty{100}{\AU}}{d} \right) \, .
\end{equation} 

As the emitted GWs remove energy, the orbital radius decays and the GW frequency increases.
The GW frequency at early times is given by~\cite{Franciolini:2022htd}
\begin{align}
f_{\rm GW}(t) &= f_{0, \rm GW}\left( 1-\frac{t}{t_c} \right)^{-3/8}\,. \label{eq:omega_t}
\end{align}
Here, $f_{0, \rm GW}$ is a reference GW frequency at $t = 0$, when the PBHs are well-separated and the point-like approximation holds. Then $f_{\rm GW}$ diverges within a finite coalescence time
\begin{align}
t_c & \approx \qty{e-7}{\second} \, \left(\frac{\qty{e-7}{\msol}}{m} \right)^{5/3} \left( \frac{\qty{1}{\GHz}}{f_{0,\rm GW}} \right)^{8/3} \,. \label{eq:timescale}
\end{align}
This sets the characteristic timescale over which the frequency changes by $\mathcal{O}(1)$ near $f_{0,{\rm GW}}$. 

In reality, $f_{\rm GW}(t)$ is bounded: the binaries eventually reach the innermost stable circular orbit with $R_{\rm ISCO} = 12\, G_N m$, at which point the GW frequency is $f_{\rm ISCO}\approx \qty{2.2}{\GHz} \, (\qty{e-6}{\msol}/m)$.
This is the maximum frequency before the point-like approximation breaks down and the PBHs enter the merger and ringdown phases, which we do not consider.

Transient radio sources are characterized by their fluence $\mathcal{F}$, the energy received per unit frequency per area, integrated over the burst. For a merger at $d$ observed in a band $[f_{\rm lo}, f_{\rm hi}]$, we compute the observed bandwidth-averaged fluence as
\begin{equation}
\mathcal{F} \approx \frac{1}{4\pi d^2} \frac{1}{\Delta f} \int_{f_{\rm lo}}^{f_{\rm max}} df_{\rm GW}\, P_{h \to \gamma} \frac{d E_{\rm GW}}{df_{\rm GW}} \, 
\end{equation}
with $\Delta f \equiv f_{\rm hi} - f_{\rm lo}$. The upper limit $f_{\rm max}= \min\{f_{\rm ISCO}, f_{\rm hi}\}$ truncates the integral at $f_{\rm ISCO}$ when it lies inside the band, so only emission for which the inspiral formula applies is counted (we set $\mathcal{F} = 0$ when $f_{\rm ISCO} < f_{\rm lo}$).
In the limit where $B$, $\ell_c$, and $n_e$ are constant along the path and $\ell_{\rm osc} \gg \ell_c$, the fluence is approximately
\begin{alignat}{2}
\mathcal{F} &\sim && \,  \qty{0.4}{\jansky\milli\second} \left( \frac{B}{\qty{5}{\micro\gauss}} \right)^2\left( \frac{\qty{400}{\mega\hertz}}{\Delta f} \right)  \left( \frac{\qty{100}{\AU}}{d} \right)\nonumber \\
& && \times \left( \frac{\ell_c}{\qty{0.1}{\AU}} \right) \left( \frac{m}{\qty{e-6}{\msol}} \right)^{5/3} \left( \frac{f_{\rm lo}}{\qty{400}{\mega\hertz}} \right)^{2/3} \! \!\!\! .
\end{alignat}

\paragraph*{\textbf{Sensitivity of Radio Telescopes to PBH Mergers.}}
We now estimate the sensitivity of two existing radio telescopes to PBH mergers: the Canadian Hydrogen Intensity Mapping Experiment (CHIME, 400--\qty{800}{\mega\hertz})~\cite{CHIME:2022dwe} and the Five-hundred-meter Aperture Spherical Telescope (FAST, 1.05--\qty{1.45}{\giga\hertz})~\cite{zotero-item-3161,zotero-item-3149,zotero-item-3151}. 
Each telescope has detected hundreds of fast radio bursts (FRBs)~\cite{CHIMEFRB:2021srp}, which are bright, extremely short-lived radio sources with typical fluences $\mathcal{F} \sim \qty{1}{\jansky\milli\second}$.
As such, we adopt a detection threshold for radio emission from PBH mergers of $\mathcal{F} > \qty{0.5}{\jansky\milli\second}$, noting that FRBs with $\mathcal{F} \sim \qty{0.05}{\jansky\milli\second}$ are routinely detected~\cite{CHIMEFRB:2021srp}.

\begin{figure}[t]
    \centering
    \includegraphics[width=\linewidth]{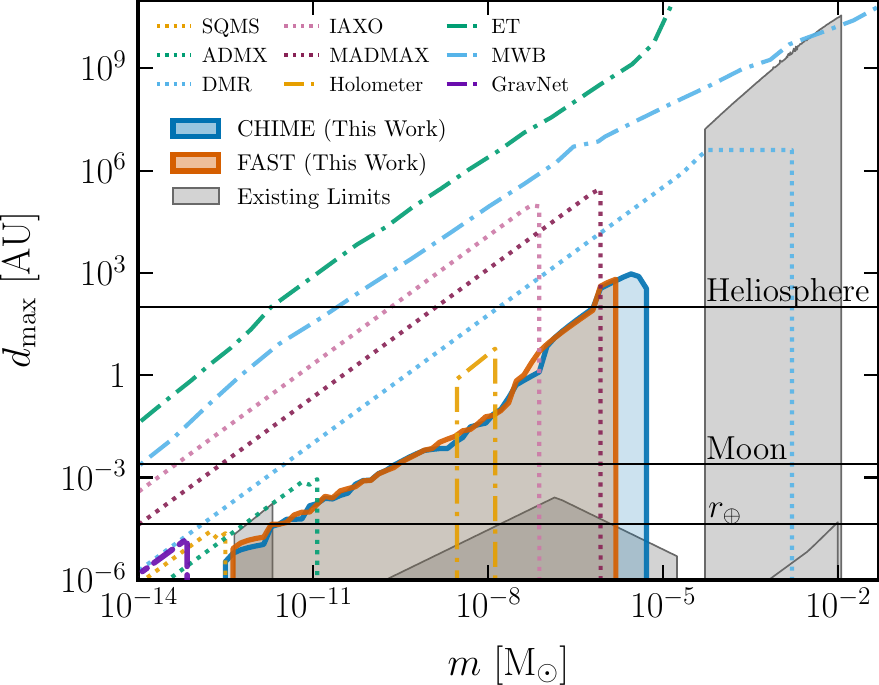}
    \caption{Maximum detectable distance $d_{\rm max}$ for equal-mass PBH mergers vs.\ PBH mass $m$. Solid/filled curves: CHIME (blue) and FAST (orange) (this work). Grey: existing limits from LIGO~\cite{Miller:2021knj,Miller:2024fpo}, ADMX-SLIC~\cite{Crisosto:2019fcj}, ABRACADABRA~\cite{Salemi:2021gck}, and the Holometer~\cite{Holometer:2016qoh}. Dotted: proposed axion experiments; dash-dotted: proposed GW detectors~\cite{Ringwald:2020ist,Aggarwal:2020olq,Berlin:2021txa,Domcke:2022rgu,Franciolini:2022htd,Domcke:2023bat,Domcke:2024mfu}. Horizontal lines mark reference distances from Earth's surface.}
    \label{fig:pbh_sens}
\end{figure}

Since $\mathcal{F}$ depends on PBH mass and distance, for each mass we solve for the maximum distance $d$ at which $\mathcal{F} > \qty{0.5}{\jansky\milli\second}$, and the result for each telescope is shown in Fig.~\ref{fig:pbh_sens}.
CHIME and FAST have comparable sensitivity to PBH mergers for equal-mass mergers with $\qty{e-13}{\msol} \lesssim m \lesssim \qty{e-6}{\msol}$. 
At the high end of this mass range, the telescopes are sensitive to PBH mergers at $d\lesssim \qty{1000}{\AU}$, with sensitivity weakening at lower masses where strains are smaller. 
CHIME has slightly greater sensitivity at high masses since it observes at lower frequencies, where the binary spends more time emitting in-band.
For large $m$, $f_{\rm ISCO}$ drops below the band, giving a hard high-mass cutoff. 

\begin{figure}[t]
    \centering
    \includegraphics[width=\linewidth]{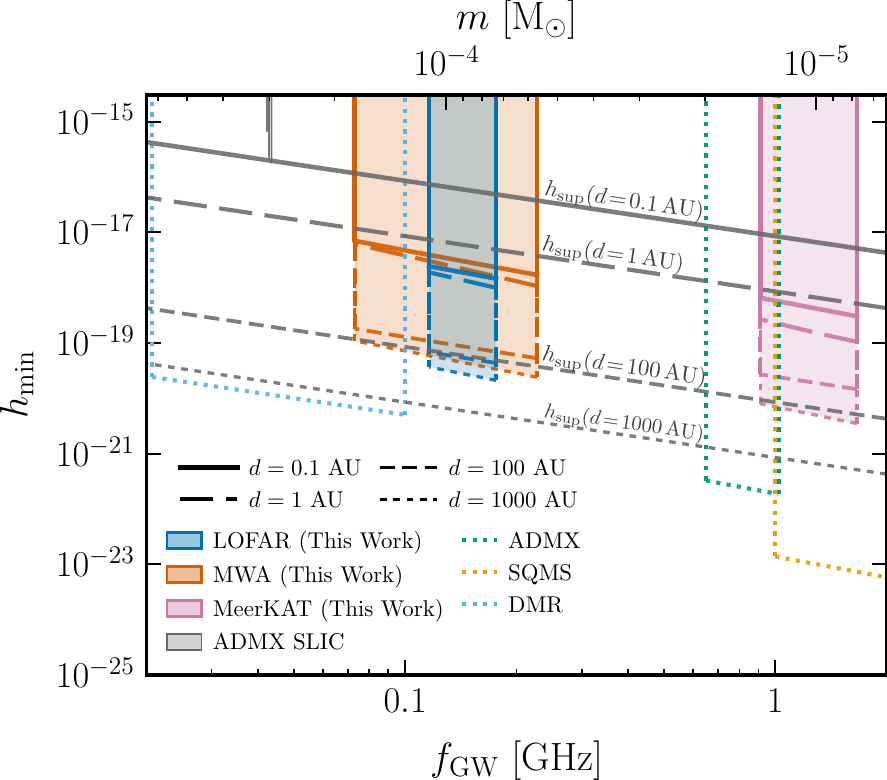}
    \caption{Minimum detectable strain $h_{\rm min}$ at Earth for a coherent HFGW source from LOFAR (blue), MWA (orange), and MeerKAT (pink); line styles indicate source distance. Dotted: projected sensitivities of proposed experiments \cite{Berlin:2021txa,Domcke:2022rgu,Domcke:2023bat}; grey region: recast ADMX-SLIC limits~\cite{Crisosto:2019fcj,Domcke:2023bat}. Grey: predicted PBH-superradiance strain at several distances assuming the parameters in Eq.~\eqref{eq:sup_strain}; the radio signal is detectable when $h_{\rm sup}$ is larger than $h_{\rm min}$ at a given $d$. The upper axis shows the required PBH mass for the signal for $\alpha=0.2$ and $\Delta \chi=0.5$.}
    \label{fig:super_sens}
\end{figure}

PBH mergers have distinct radio signatures that set them apart from other transient sources.
With sufficiently good time resolution, the low-to-high frequency chirp would lead to a highly distinctive negative dispersion; otherwise, mergers on the high-mass end would appear as broadband, nonrepeating, point-like bursts with negligible DM and no counterpart across the electromagnetic spectrum.
Such a signature would be highly unusual for an astrophysical source; definitively distinguishing it from radio-frequency interference may, however, pose a challenge, although the point-like nature and the possibility of detection at multiple telescopes could mitigate this issue.
Nevertheless, nearby mergers can be extremely bright, and should be detectable regardless.

We compare our results in Fig.~\ref{fig:pbh_sens} to existing limits (grey) from LIGO~\cite{Miller:2021knj,Miller:2024fpo}, the Holometer~\cite{Holometer:2016qoh}, ADMX-SLIC~\cite{Crisosto:2019fcj}, and ABRACADABRA~\cite{Salemi:2021gck}.
The strongest current limits come from a search for the slowly-evolving early-inspiral emission in LIGO data.
We also show forecasts~\cite{Domcke:2024mfu} for the Einstein Telescope (ET)~\cite{Hild:2008abc,Punturo:2010zz,Hild:2010id} and a magnetic Weber bar (MWB)~\cite{Berlin:2023grv}, which may probe the early-inspiral GWs or the model-dependent GW memory from PBH mergers~\cite{Gasparotto:2025wok,Berlin:2026che}.
For ET and MWB, we convert the forecasted PBH abundance sensitivities of Ref.~\cite{Gasparotto:2025wok} into limits on $d_{\rm max}$.
The Holometer, a Michelson interferometer, searched for the monochromatic GW signal from PBH mergers (grey); a forecasted search for the chirp signal would significantly improve its sensitivity (dash-dotted yellow)~\cite{Holometer:2016qoh}.

The ADMX-SLIC and ABRACADABRA limits are recast from axion limits following Refs.~\cite{Berlin:2021txa,Domcke:2022rgu,Domcke:2023bat}. Reported limits on the axion-photon coupling $g_{a\gamma \gamma}$~\cite{Crisosto:2019fcj,Salemi:2021gck} correspond to a minimum detectable axion-induced magnetic flux $\Phi_{B,a}$, which can be mapped onto a minimum detectable GW-induced magnetic flux $\Phi_{B,h} \propto h$ via $\Phi_{B,h} = \mathcal{R}_c \Phi_{B,a}$~\cite{Domcke:2022rgu,Domcke:2023bat}. For a resonant cavity like ADMX-SLIC, the coherence ratio is~\cite{Domcke:2023bat}
\begin{equation}
    \label{eq:RC}
    \mathcal{R}_c = \left(\frac{T_m}{\tau_h}\right)^{1/4} \left(\frac{Q_a}{Q_h}\right)^{1/4} A \, ,
\end{equation}
with axion-signal quality factor $Q_a \sim 10^6$, cavity quality factor $Q_r$, per-mass scan time $T_m$, GW quality factor $Q_h$, and coherence time $\tau_h \equiv Q_h/f_{\rm GW}$.
Here, $A=1$ if $Q_r < Q_a, Q_h$ and $A=Q_r/ Q_h$ if $Q_h < Q_r < Q_a$ \cite{Domcke:2023bat}. 
For a broadband experiment like ABRACADABRA, $\mathcal{R}_c = \left(Q_a/Q_h\right)^{1/4}$ \cite{Domcke:2022rgu}. 
Explicit expressions for $\Phi_{B,a}$ and $\Phi_{B,h}$ for different detector geometries are given in Ref.~\cite{Domcke:2023bat}. 
For PBH mergers~\cite{Domcke:2022rgu,Domcke:2023bat},
\begin{equation}
Q_h \sim \frac{f_{\rm GW}^2}{\dot{f}_{\rm GW}} \approx 2 \times 10^6 \left(\frac{\qty{1}{\giga\hertz}}{f_{\rm GW}}\right)^{5/3}\left(\frac{\qty{e-9}{\msol}}{m}\right)^{5/3} \, . \label{eq:Q}
\end{equation}
For mergers of heavy PBHs, $Q_h$ is generally much less than $Q_r$ and $Q_a$. In turn, $\mathcal{R}_c \gg 1$ and the signal becomes more difficult to detect.
While lighter PBHs would lead to a larger $Q_h$, the strain from such mergers scales unfavorably as $m^{5/3}$. 
We compute $Q_h$ and $\tau_h$ from Eq.~\eqref{eq:Q} and solve for the $d_{\rm max}$ at which the PBH strain in Eq.~\eqref{eq:pbh_strain} matches the recast limit on $h$.

Fig.~\ref{fig:pbh_sens} also shows forecasts for several proposed terrestrial experiments. 
For DMRadio, we use the same method described above, assuming a \qty{100}{\meter\cubed} volume and a different pickup-loop geometry than would be used in the main axion experiment~\cite{Domcke:2022rgu,Domcke:2023bat}. 
ADMX, SQMS cavities, and GravNet are sensitive to strains~\cite{Berlin:2021txa,Franciolini:2022htd,Domcke:2022rgu}
\begin{alignat}{2}
    \label{eq:cavity_sens}
    h &\sim && \, 10^{-22} \left(\frac{\qty{0.1}{\meter\cubed}}{V_{\rm cav}}\right)^{5/6}\left(\frac{10^5}{Q_r}\right)^{1/2} \left(\frac{0.1}{\eta_n}\right) \left(\frac{T_{\rm sys}}{\qty{1}{\kelvin}}\right)^{1/2} \nonumber \\
    & && \times  \left(\frac{\qty{8}{\tesla}}{B}\right) \left(\frac{\qty{1}{\giga\hertz}}{f_{\rm GW}}\right)^{3/2} \left(\frac{\Delta f}{\qty{10}{\kilo\hertz}}\right)^{1/4}\left(\frac{\qty{1}{\minute}}{t_{\rm obs}}\right)^{1/4} \, ,
\end{alignat}
where $V_{\rm cav}$ is the cavity volume, $\eta_n$ the cavity-GW coupling, $T_{\rm sys}$ the system temperature, $B$ the magnetic field, $\Delta f$ the bandwidth, and $t_{\rm obs} = t_{\rm lo} [ 1- ( f_{\rm lo}/f_{\rm max})^{8/3}]$, with $t_{\rm lo}=t_c$ from Eq.~\eqref{eq:timescale} at $f_{0,{\rm GW}} = f_{\rm lo}$.
We adopt the experimental parameters from Ref.~\cite{Franciolini:2022htd} for ADMX and an SQMS-developed cavity, and Ref.~\cite{Schmieden:2023fzn} for GravNet.
Eq.~\eqref{eq:cavity_sens} assumes $Q_h > Q_r$; in the opposite regime where $Q_h < Q_r$, we set $d_{\rm max}=0$ for simplicity following Ref.~\cite{Franciolini:2022htd}, although the reduction in sensitivity should be more gradual following Eq.~\eqref{eq:RC}.

For IAXO and MADMAX, the parametric strain sensitivity is \cite{Ringwald:2020ist,Franciolini:2022htd}
\begin{alignat}{2}
    \label{eq:spd_sens}
    h & \sim && \,  10^{-24} \left(\frac{\qty{1}{\meter}}{L}\right)\left(\frac{\qty{1}{\meter\squared}}{A}\right)^{1/2} \nonumber \\
    & && \times \left(\frac{\qty{1}{\tesla}}{B}\right) \left(\frac{\qty{100}{\giga\hertz}}{\Delta f}\right)^{1/2}\left(\frac{\qty{1}{\year}}{t_{\rm obs}}\right)^{1/4}\, ,
\end{alignat}
where $L$ and $A$ are the length and area of the conversion region. 
Note that the IAXO projection assumes a radio receiver rather than the X-ray detector of the main proposal~\cite{Ringwald:2020ist,Aggarwal:2020olq}.

\paragraph*{\textbf{Black Hole Superradiance.}}

Longer-lasting, highly monochromatic HFGW sources may also exist, e.g.\ GW emission from ultralight boson clouds formed by superradiance around a PBH~\cite{Arvanitaki:2010sy}.
The peak strain of the GW emitted (observed at a distance $d$ from the source) is \cite{Arvanitaki:2014wva,Arvanitaki:2016qwi,Brito:2015oca,Aggarwal:2020umq} 
\begin{equation}
\label{eq:sup_strain}
h_{\rm sup} \sim 10^{-18} \left(\frac{\alpha}{0.2}\right)^7 \left(\frac{\Delta \chi}{0.5}\right) \left(\frac{\qty{100}{\AU}}{d}
\right)\left(\frac{m}{\qty{e-3}{\msol}}\right)
\, ,
\end{equation}
assuming an $\ell= m_\ell = 1$ gravitational atom state. Here, $\alpha \equiv G_N m \, m_b$ for boson mass $m_b$ and $\Delta\chi$ is the change in dimensionless BH spin ($0 \leq \chi < 1$) over the superradiance process. 
The emission lasts for~\cite{Franciolini:2022htd,Brito:2015oca}
\begin{equation}
\tau \sim \qty{4e-4}{\year}\,\left( \frac{m}{\qty{e-3}{\msol}} \right) \left( \frac{\alpha}{0.2} \right)^{-15} \left( \frac{\Delta \chi}{0.5} \right)^{-1} \,.
\end{equation} 
For this gravitational atom mode, superradiance requires $\alpha \leq \chi [1 + \sqrt{1 - \chi^2}]^{-1}/2$~\cite{Aggarwal:2020umq}, and the steep $\alpha$-dependence of $h_{\rm sup}$ and $\tau$ means detectable strains arise only when $\alpha \gtrsim \mathcal{O}(0.1)$ or $m \sim (G_N m_b)^{-1}$, fixing the GW frequency to $f_{\rm GW} = m_b / \pi \sim \qty{100}{\MHz} (\qty{e-4}{\msol} / m)$.
This therefore requires a primordial black hole, and it is unclear whether superradiant spindown would have already completed, or whether superradiance occurs at all given the low initial spins of PBHs~\cite{DeLuca:2019buf,Franciolini:2022htd}.
Nevertheless, this signal would appear as a bright, monochromatic radio source, with detectable signals preferring shorter lifetimes.

We characterize detectability of this signal by the bandwidth-averaged spectral flux density $\langle S_{\rm EM}\rangle$ rather than fluence.
For a monochromatic source,
\begin{equation}
    \langle S_{\rm EM}\rangle = P_{h\to \gamma} \frac{\Phi_{\rm GW}}{\Delta f} = \frac{h^2 \omega_{\rm GW}^2}{32 \pi G_N \Delta f} P_{h \to \gamma} \, ,
\end{equation}
using the flux of a monochromatic plane GW~\cite{Maggiore:2007abc}. Numerically, 
\begin{alignat*}{2}
    \langle S_{\rm EM} \rangle &= && \, \qty{0.1}{\jansky}\left( \frac{h}{10^{-19}}\right)^2  \nonumber \\
    & && \times \left(\frac{f_{\rm GW}}{\qty{100}{\mega\hertz}}\right)^2  \left(\frac{\qty{100}{\mega\hertz}}{\Delta f}\right) \left(\frac{P_{h \to \gamma}}{10^{-32}}\right) \,.
\end{alignat*}
We require $\langle S_{\rm EM}\rangle > \qty{0.1}{\jansky}$ for detectability, a conservative choice since the faintest observed radio sources lie below \qty{0.1}{\milli\jansky}.

In Fig.~\ref{fig:super_sens}, we show the strain sensitivity $h_{\rm min}$ at Earth for LOFAR (115--\qty{177}{\mega\hertz})~\cite{Mandal:2020zor,Shimwell:2025tui}, MeerKAT (900--\qty{1670}{\mega\hertz})~\cite{Mutale:2026}, and MWA (72--\qty{231}{\mega\hertz})~\cite{Hurley:2017}. 
Since the magnetic field properties vary as a function of distance, $h_{\rm min}$ depends on the distance of a source from Earth.
For nearby sources, $P_{h\to \gamma}$ is small and the sources must be intrinsically bright, leading to a larger $h_{\rm min}$.
For more distant sources, $P_{h \to \gamma}$ is larger, and intrinsically fainter sources can be detected.
However, $P_{h\to\gamma}(d)$ begins to saturate beyond $\sim \qty{100}{\AU}$ (Fig.~\ref{fig:lengths_and_P}), and $h_{\rm min}$ saturates for distances $d \gtrsim \qty{300}{\AU}$.
Comparing to the predicted  $h_{\rm sup}$ (grey) for the parameter choices in Eq.~\eqref{eq:sup_strain}, PBH superradiance is detectable where $h_{\rm sup} > h_{\rm min}$, i.e.\ for $d \lesssim \qty{100}{\AU}$, though again this depends strongly on $\alpha$. 

We also plot the same proposed experiments as in Fig.~\ref{fig:pbh_sens} and recast ADMX-SLIC limits~\cite{Domcke:2022rgu,Domcke:2023bat}, adopting integration times of \qty{2}{\minute} (ADMX) and 1 day (SQMS), and $Q_h=10^3$ (DMRadio)~\cite{Domcke:2022rgu}. These experiments can outperform radio telescopes for monochromatic sources if they are tuned to the right frequency and can integrate for a long time, but radio telescopes still extend the search to new parameter space, and archival data may enable an immediate search.

\paragraph*{\textbf{Discussion.}}
While the estimated merger rate of $m \sim \qty{3e-6}{\msol}$ PBHs within a volume of radius $\sim \qty{1000}{\AU}$ from Earth is extremely low ($\sim 10^{-20}$ mergers per year following the methods of Ref.~\cite{Franciolini:2022htd}), Fig.~\ref{fig:pbh_sens} more broadly shows that radio telescopes can significantly outperform existing search efforts for generic HFGW sources. 
Given that any realistically detectable source of HFGWs in the GHz regime must lie within the Solar System on model-independent grounds~\cite{Berlin:2026che}, radio telescopes may already be capable of completely ruling out HFGW scenarios that can be practically discovered. 
At the very least, radio telescopes should be viewed as one of the most pragmatic avenues to search for these rare events.

A dedicated search for PBH-merger radio bursts or highly monochromatic radio point sources could significantly improve on our estimated sensitivity.
Additionally, planned facilities hold even greater promise.
CHORD~\cite{Vanderlinde:2019tjta}, under construction, will detect fainter, shorter transients over a larger bandwidth than CHIME or FAST, improving its sensitivity to PBH mergers.
SKA (also under construction) \cite{Braun:2019gdo,Weltman:2018zrl} will have exceptional sensitivity to faint radio point sources and will therefore be highly sensitive to monochromatic HFGWs as well.

\paragraph*{\textbf{Acknowledgements.}}
We thank Asher Berlin, Robert Pascua, Nicholas Rodd, and Sophia Rubens for useful conversations. 
The work in this paper makes extensive use of the \textsc{numpy}~\cite{Harris:2020xlr}, \textsc{scipy}~\cite{Virtanen:2019joe}, \textsc{matplotlib}~\cite{Hunter:2007ouj}, and \textsc{astropy}~\cite{Astropy:2013muo,Astropy:2018wqo,Astropy:2022ucr} packages.
EB and HL are supported by the U.S. Department of Energy under grant DE-SC0026297 and the Cecile K. Dalton Career Development Professorship, endowed by Boston University trustee Nathaniel Dalton and Amy Gottlieb Dalton. 

\bibliography{sources}

\end{document}